\documentclass[12pt]{article}
\usepackage[english]{babel}
\usepackage[latin1]{inputenc}
\usepackage{amsmath, amsthm, amssymb}
\usepackage[pdftex]{graphicx}
\usepackage{amssymb}
\usepackage{latexsym}
\usepackage{amstext}
\usepackage{epstopdf}
\usepackage{floatflt}
\usepackage{hyperref}
\usepackage{enumitem}
\usepackage{multirow}
\usepackage{pdflscape}
\usepackage{makecell}

\newcommand{\aaa}{\mathfrak{a}}
\newcommand{\ad}{\aaa_2}

\newcommand{\adc}{\aaa_2^c}

\newcommand{\adf}{\aaa_2^f}

\newcommand{\as}{\alpha}
\newcommand{\asf}[1]{\alpha_{F\, #1}}

\newcommand{\ba}{\begin{array}}

\newcommand{\be}{\begin{equation}}
\newcommand{\bea}{\begin{equation}\begin{array}}
\newcommand{\beal}{\begin{aligned}}
\newcommand{\beas}{\begin{equation*}\begin{array}}
\newcommand{\bef}{\begin{flalign}}
\newcommand{\befs}{\begin{flalign*}}
\newcommand{\bes}{\begin{equation*}}

\newcommand{\bit}{\begin{itemize}}

\newcommand{\dotto}{\mathbf{d_8}}

\newcommand{\dtre}{\mathbf{d_3}}
\newcommand{\ea}{\end{array}}

\newcommand{\eal}{\end{aligned}}

\newcommand{\ee}{\end{equation}}
\newcommand{\eea}{\end{array}\end{equation}}
\newcommand{\eeas}{\end{array}\end{equation*}}
\newcommand{\eef}{\end{flalign}}
\newcommand{\eefs}{\end{flalign*}}
\newcommand{\ees}{\end{equation*}}

\newcommand{\eit}{\end{itemize}}

\newcommand{\eo}{\mathbf{e_8}}

\newcommand{\es}{\mathbf{e_6}}

\newcommand{\Kass}{\mathbb K}

\newcommand{\lam}{\lambda}

\newcommand{\nin}{\noindent}

\newcommand{\Perm}{\mathbb P}

\numberwithin{equation}{section}
\begin{document}

\begin{center}
{\LARGE Vertex operators for an expanding universe}\footnote{The following article will be published by AIP Publishing, Conference Proceedings of {\it Symmetries and Order: Algebraic Methods in Many Body Systems} in honor of prof. Francesco Iachello, Yale October 5-6 2018}
\vspace{5mm}

{\Large Piero Truini}\\
I.N.F.N. via Dodecaneso 33, 16146 Genova - Italy\\
QGR 101 S. Topanga Canyon Rd. 1159 Los Angeles, CA 90290 - USA

\vspace{5mm}

truini@ge.infn.it
\end{center}

\begin{abstract}
I  am presenting a quantum model for the universe at its early stages that includes a mechanism for the creation of space, starting from an initial quantum state and driven by $\eo$ interactions.
\end{abstract}

\section{INTRODUCTION}

Physics is fascinating! It is the science that delves more deeply into the study of nature. By doing so, it reveals a world that goes far beyond common intuitions, opening a Pandora's vase of new problems and mysteries.
It is amazing that gravity, the most intuitive force in our daily lives, carries the hardest challenge in the current frontier of physics.\\
Gravitational physics spans from cosmology and astrophysics on the large scale to elementary-particle quantum physics on the small scale.
General relativity, together with the observation that the universe is expanding, implies that the universe began in a big bang.
The two regions of large and small scales, so far apart in the present observations, merge into a single one in the early universe, when the particles that originated what we observe today were compressed within a volume of few Planck lengths per size. The energy in such a compressed universe, the Planck energy of $10^{16}\ Tev$, is a thousand trillion times greater than the energy at LHC. At the enormous densities and energy of the big bang era all forces are unified, and spacetime itself is a mysterious object, maybe discrete or quantized, maybe not even primordial. General relativity and cosmological observations make physicists believe that space itself is expanding.\\
The current challenge of fundamental physics is to have a consistent model for the earliest moments of the universe that unifies gravity - hence the concept of spacetime - with the other forces known in nature. The quest is for a quantum theory of gravity that includes the notion of a quantum initial state of the universe. Many theoretical physicists are making a big effort in trying to solve the mysteries of quantum gravity, by extending  their knowledge on the most advanced theories to the highest density region. Quantum cosmology, however, faces  a problem that is profoundly different from those so far encountered  in physics. This is precisely the  need for a theory on the initial condition of the universe.\\
This paper moves a few  steps in this direction. What I am presenting here tends to be very elementary and basic through the use the Occam's razor. The aim is  to find a model for the expansion of the universe from an initial quantum state, based on few fundamental principles and observations from quantum physics. The model is based on the Lie algebra $\eo$, whose generators act as vertex operators on a discrete space that is being built up step by step by $\eo$ driven interactions.
Vertex operator algebras were introduced originally in String Theory and conformal field theory in 2 dimensions, \cite{belavin}-\nocite{verl}\cite{huang}, to describe certain types of interactions between different strings, localized at vertices, analogous to the Feynman diagrams' vertices. Mathematically the underlying concept of a vertex algebra was introduced by Borcherds, \cite{borch}-\nocite{frenk}\cite{kacvoa}, in order to prove the {\it monstrous moonshine} conjecture, \cite{conway1}. I am using here the term {\it vertex operator} in a broader sense, referring to interaction operators that look like generators of a Lie algebra, $\eo$ in particular, but whose product depends upon parameters related to the spacetime creation and expansion.
The goal is to produce a consistent model with calculable quantities that can be derived from a density matrix - like the partition function, the mean energy and the von Neumann entropy - when a few Plank units of time have elapsed since the initial quantum state.\\
Like every physicists, I never give up the hope hidden at the bottom of  Pandora's vase.

\section{{\LARGE $\eo$} IN A PHYSICAL PERSPECTIVE}

Exceptional groups and algebras and the underlying non-associative algebra of the octonions have attracted the attention of many theoretical Physicists since the pioneering work of F. G\"{u}rsey, \cite{gursey} \cite{bied1}. G\"{u}rsey noted that specializing one of the seven nonscalar Cayley units (to play the role of the imaginary unit) automatically achieves a rationale for $SU(3)^{color}$. At that time $\es$ was considered one of the best candidate for a theory of Grand Unification.\\
If we look inside $\eo$ we see four orthogonal $\mathbf {su(3)}$'s - $\ad$'s in their complex form. Beside $\adc$, for color we see
$\adf$ that can be associated to flavor degrees of freedom and two more, \cite{pt1} \cite{T-2}. It is natural to associate these two $\ad$'s to degrees of freedom related to gravity. They appear on equal footing as color and flavor symmetries in the realm of a quantum gravity theory.\\
It seems therefore natural to think of $\eo$ as a model for a theory that includes spacetime, whose classical concept needs to be reviewed in the light of  the incompatibility of quantum mechanics and general relativity at the Planck scale. This incompatibility at small scales makes it impossible to test, in particular, if spacetime is a continuum.\\
I do believe, following the Occam's principle, that a genuine new theory should be based on the least possible assumptions on the basic laws of Physics. As Wheeler said:\\
\nin {\it It is my opinion that everything must be based on a simple idea. And it is my opinion that this idea, once we have finally discovered it,
	will be so compelling, so beautiful, that we will say to one another, yes, how could it have been any different.}   J.A. Wheeler\\
An important feature of spacetime is that it is {\em dynamical} and related to matter, as Einstein taught us in his theory of general relativity. The big bang, for instance, is not a blast in empty space. Physicists do not think there was a space and the big bang happened in it: big bang was a blast of spacetime itself, as well as matter and energy. Therefore from the idea that spacetime is dynamical and that it is being created I deduce that:\\
\nin {\it Basic Principle: There is no way of defining spacetime without a preliminary concept of interaction}.\\
Stated differently, a universe of non-interacting particles has no spacetime: there is no physical quantity that can relate one particle to another. We are used to start from spacetime because our point of view is that of an observer, who measures things in spacetime. However there are no external observers with clocks in the cosmological context of the big bang era. The basic principle implies that one has to start from a model of interactions, consistent with the present observations, and deduce from  it what spacetime is. This is the way I look at $\eo$ - and its finite algebraic extensions called exceptional periodicity, \cite{trm}.\\
All fundamental interactions look similar at short distances. Their basic structure is very simple: it involves only three entities, like the product in an algebra. The first step in my approach is to define objects and {\it elementary interactions}, with the hypothesis in mind, similar to the Bethe \textit{Ansatz} in the Heisenberg model, that every interaction is made of elementary interactions. This hypothesis gives the interactions a tree structure, thus opening the way for a description of scattering amplitudes in terms of associahedra or rather permutahedra as I will discuss below, \cite{marni}-\nocite{mike}\nocite{nima1}\nocite{miz}\nocite{nima2}\nocite{stash1}\nocite{stash2}\nocite{tonks}\cite{loday}.\\
\begin{figure}
  \centerline{%
  \includegraphics
   [width=.6\textwidth]{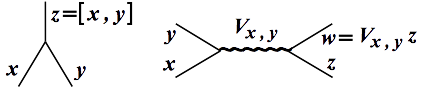}}
 \caption{Building blocks of the interactions and an elementary scattering process.}
 \label{int1}
\end{figure}
The proposal is to start from $\eo$, or extensions of it, and state that the algebra rules determine the building blocks of interactions. An elementary - or fundamental - interaction may be defined as the interaction between $x$ and $y$ in $\eo$ to produce the outcome $z$ in $\eo$. It is represented by $(x,y ->z) \leftrightarrow [x,y] = z$, see figure \ref{int1}. Jordan pairs, \cite{loos}, perfectly fit into this picture as matter particle-antiparticle pairs, the quarks in the octonionic Jordan pair triple and the leptons in the complex one inside $\es$, while  an elementary scattering is represented by the triple product $V_{x_i^\sigma , y_j^{-\sigma}}$, \cite{pt1} \cite{T-2}.

\section{EMERGENT SPACETIME}

The Occam's razor principle suggests to look at the minimal initial conditions at the big bang, namely those in which the least number of generators is taken, corresponding to roots whose linear combination generates all the roots. In the case of $\eo$ this minimal number of roots is 9: a set of simple roots, that generates all positive roots, plus 1 properly chosen negative root. Notice that limiting the number of initial generators implies a {\it symmetry breaking} with respect to the algebra of $\eo$.
A pair of opposite 3-momenta $\vec p$ and $-\vec p$ is associated to each root of the initial particles and the set of generators corresponding to these initial roots are allowed to interact among themselves. This is interpreted, having in mind locality, by the fact that the initial particles are all at the same point, even though there is no geometry, no singularity and actually no point of an \textit{a priori} space.
The assignment of opposite momenta reproduces the quantum behavior of a particle in a box where the square of the momentum has a definite value in a stationary state but not the momentum.
Particles oscillate back and forth as quantum objects (not classical ones!) due to the interaction with the border of a finite (expanding) universe: this quantum effect is the main difference between creating space and leaving in a background one. Mathematically it is the difference between working in ${\mathcal L}^2({\mathbb R})$ and ${\mathcal L}^2(a,b)$.\\
In the present algorithm all the initial roots are assigned onshell 4-momenta of massless particles, so that there is no arbitrary choice of energy. Other possibilities may be considered, like having tachyonic particles. The initial particles, see below, determine the border of the universe where they cannot interact hence be observed. On the other hand, the interaction of two non-aligned tachyons may produce massless or massive particles in the observable part of the universe.\\
Equal amplitudes are assigned to the initial generators thus completing the description the {\it initial quantum pure state}.\\
Each elementary interaction between 2 incoming particles has two effects: produce a new particle according to the commutation rules, with amplitude given by the structure constants of $\eo$, and create, with the proper amplitudes for quantum particles in a box, two new points $\pm {\vec p}/E$ apart from the point of the interaction, for each incoming particle with 4-momentum $(E,{\vec p})$. The universe thus gets into a {\it fully entangled pure state} at its early stages. It stays so at least until the first observations, as quantum mechanics teaches us. An observation, meaning a measurement by an apparatus that entangles with the observed object, may change the pure state of the universe: namely it may disentangle a very tiny part from it and eventually, after subsequent observations, turn that pure state into a mixture. Quasi-classical phenomena may thus appear in a very small part of the universe - locally, we may say. The density matrix of the universe changes a little tiny bit, some information is lost and the entropy increases.\\
Recently it has been suggested that quantum entanglement in holographic descriptions, \cite{thooft}-\nocite{suss}\cite{bousso},
plays an important role in the emergence of the classical spacetime
of general relativity, \cite{ryu}-\nocite{hrt}\cite{mald}. This raises the possibility that entanglement is indeed the defining property that controls the physics of dynamical spacetimes, \cite{nomura1} \cite{nomura2}. My model presents an alternative view of a similar concept.\\
After the first stage (or level) of interactions the outcome can be explicitly calculated and a second stage occurs and so on. One can intuitively associate a {\it cosmological discrete (quantum of) time} with each stage of interactions. At any time {\it only those generators that overlap at the same created point can interact}. This is where the model shows the vertex operator nature of the $\eo$ generators.\\
What emerge are {\it quantum fields}: each field is identified with a generator of $\eo$, that spreads in momentum space as well as in position space, like a discrete quantum wave as spacetime is created.\\
It is obvious that the spacetime emerging in the approach outlined here is dynamical, finite and discrete, being the outcome of a countable number of interactions among a finite number of objects. This is in agreement with the two cutoffs coming from our current knowledge of Physics: the background radiation temperature (finiteness) and the Planck length (discreteness). The granularity of spacetime implies that the velocity of propagation of the interaction is also discrete and finite. If the distance traveled from one level of interactions to the next one is 1 Planck length and the time interval is 1 Planck time then the maximum speed of propagation is the speed of light. The model is intrinsically relativistic, with curvature determined by all interactions. It is also quantum mechanical as it represents the quantum evolution of a quantum initial state.
This approach will lead to a finite model by construction, with the continuum limit as a macroscopic approximation.\\
All the infinities or continuities of the standard theories are not present: there are no symmetry Lie groups, just Lie algebras. This is the reason why it makes sense to look for extensions of $\eo$, like exceptional periodicity, \cite{trm}, which do not extend the Lie group $\mathbf{E_8}$ because of lack of the Jacobi identity. A good reason for extending $\eo$ is to have a greater amount of Dark matter that has no color nor flavor charge, already present in $\eo$ in the degrees of freedom of the two {\it gravity} $\ad$'s.\\
A crucial step in the model is the assignment of two opposite 3-momenta to each initial root. Three similar procedures are being considered for the projection from the eight dimensional space of the initial root vectors to 3 dimensions. One procedure goes through the Elser Sloane projection to 4 dimensions, \cite{els}, the second one is the Moody Patera projection to 3 dimensions, \cite{mp},  the third one is a projection on a $\dtre$ subalgebra root space, called {\it merkaba} projection due to the typical  resulting picture. The first two make use of the icosians. In all cases one ends up with a three dimensional space of a tetrahedron (the simple roots of an $\mathbf {a_3} \simeq \dtre$ subalgebra).\\
The boundary of the expansion are naturally determined by massless objects. By imposing $4$-momentum conservation, eventually  every root will have momenta which can be easily expressed in terms of orthogonal directions in the three dimensional space of the tetrahedron, and the discrete space on which the particles spread is a {\it generalized quasicrystal}. An approach to quasicrystals based on $\eo$ is in \cite{fang}.\\
I do not show here the actual calculations which are partly numerical and have been tested by a computer. In particular the density matrix, the von Neumann entropy, the mean energy of the universe right after the big bang can be explicitly calculated.
\subsection{FERMIONS AND BOSONS}

Given an orthonormal basis $\{ k_1,...,k_8\} \in \mathbb R^8$ I introduce the set $\Delta_F$ of simple roots of $\eo$:
\bea{l}
\asf{1} = \frac12(k_1-k_2-k_3-k_4+k_5+k_6+k_7-k_8)\\
\asf{2}= \frac12(k_1-k_2+k_3+k_4-k_5-k_6-k_7+k_8)\\
\asf{3} = \frac12(-k_1+k_2-k_3+k_4-k_5+k_6+k_7-k_8) \\
\asf{4}= \frac12(k_1+k_2-k_3-k_4+k_5-k_6-k_7+k_8)\\
\asf{5}= \frac12(-k_1-k_2+k_3+k_4+k_5-k_6+k_7-k_8) \\
\asf{6}= \frac12(k_1+k_2+k_3-k_4-k_5+k_6+k_7+k_8)\\
\asf{7}= \frac12(-k_1-k_2-k_3-k_4-k_5+k_6-k_7+k_8) \\
\asf{8}= \frac12(k_1+k_2+k_3+k_4+k_5+k_6-k_7-k_8)
\eea
Then all simple roots are fermionic, with the definition of fermionic or bosonic roots both by:
\bea{ll}
bosonic: & \pm k_i \pm k_j \ ; \ 1\le i < j \le 8\\
fermionic: & \frac12(\pm k_1\pm k_2\pm k_3\pm k_4\pm k_5\pm k_6\pm k_7\pm k_8)
\eea
and by their height: even for bosons, odd for fermions - if $\as=\sum_{i=1}^8{\lam_i\asf{i}}$ then $ht(\as):=\sum{\lam_i}$ is its height. Bosons are $\dotto$ generators, fermions are  $\dotto$ spinors.
The negative root $\as_9$ that completes the set of roots associated to the initial generators - hence the initial quantum state - can be taken as the root of lowest height (-29), which is also fermionic:
\be
\asf{9} = -(2 \as_1 + 3 \as_2 + 4 \as_3 + 5 \as_4 + 6 \as_5 + 3 \as_6 + 4 \as_7 + 2 \as_8)
\ee

\subsection{COALGEBRA OF SPACE EXPANSION, ASSOCIAHEDRA, PERMUTAHEDRA, GRAVITAHEDRA}
\begin{figure}
\centerline{%
\includegraphics[scale=0.28]{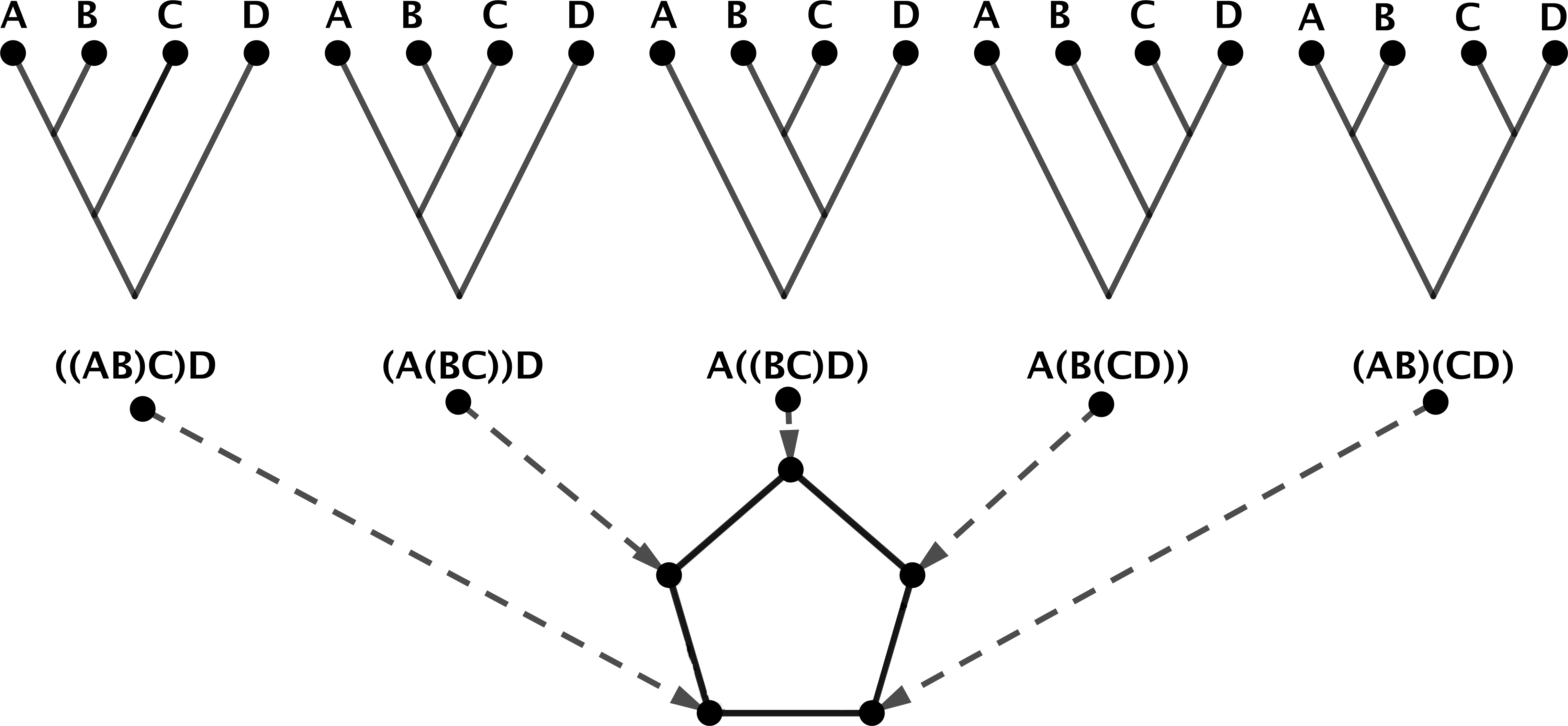}}
\caption{Associahedron $\Kass_4$. Adjacent vertices $(st)u \to s(tu)$, for subwords $s,t,u$} \label{f:Ktrees}
\end{figure}
\begin{figure}
\centerline{%
\includegraphics[scale=0.9]{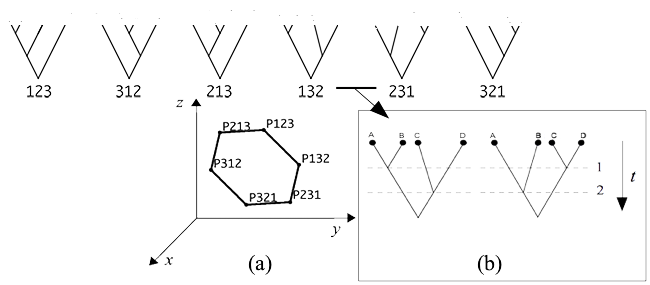}}
\caption{Permutahedron $\Perm_3$. Interaction of 4 particles.} \label{f:Ptrees}
\end{figure}

Consider the standard coalgebra associated to the Universal enveloping algebra of $\eo$, with co-product $\Delta: x \to x\otimes 1+1\otimes x$, co-unity $\epsilon: x\to 0$, antipode $S: x \to -x$ for all generators $x\in \eo$, $\Delta(1)= 1\otimes 1$, $\epsilon(1)= 1$, $S(1)=1$.\\
I introduce the notation $\Delta^S:=(Id\otimes S)\Delta$. Then the expansion of spacetime can be mathematically described by $\Delta^S$, hence by the coalgebra part of a bialgebra based on $\eo$.\\
The tree structure of the interactions allows for a description of scattering amplitudes in terms of associahedra, with structure constants attached to each vertex, see figure \ref{f:Ktrees} for the interaction of 4 particles, producing the associahedron $\Kass_4$. A vertex is interpreted as an interaction with universal time flowing from top to bottom in the trees of figure \ref{f:Ktrees}.\\
However if one includes the {\it gravitational effect} of space expansion,  one should describe the interactions through premutahedra $\Perm_{n-1}$ rather than associahedra, see figure \ref{f:Ptrees} for the interaction of 4 particles, producing the permutahedron $\Perm_3$.
 The two trees in Figure \ref{f:Ptrees} (b) are different due to the spreading of particles in space, because the same interactions occur at different times (represented by the horizontal lines).\\
A complete graphical description of the interactions, including the spacetime effects, hence gravity, can be quite complicated and needs a deep study. A research program with this goal has initiated  and the name {\it gravitahedra} has been coined for the polytopes that will eventually, and hopefully, describe such interactions.

\section{CONCLUSION}

A quantum theory of the creation of spacetime starting from a quantum state as cosmological boundary condition  is a necessity for a fundamental theory of quantum gravity. I have presented the general framework of a workable model that can be applied with confidence to the quantum era of the first cosmic evolution, based on the algebra $\eo$ and fulfilling this requirement. The model can accommodate the degrees of freedom of the particles we know, plus dark matter, plus fermions and bosons in the same unifying structure.\\
Many physical properties have still to be verified and/or fulfilled, like the charges and handedness of the Standard model particles, the Pauli principle, the proton decay,  the confinement of quarks, the attractive nature of gravity on the large scale.\\
The general framework of the model leaves however a great freedom of choice. This is as a benefit for those who believe this is a promising approach and want to explore it.

\section{ACKNOWLEDGMENTS} I wish to thank Franco Iachello for allowing me to contribute to the proceedings of the conference held at Yale in his honor. I am very proud to have been one of his collaborators, and I miss the time spent at Yale working with him. I also wish to thank Franco and Sultan Catto, for the pleasant conversations about physics we recently had in Prague. I thank Sultan for sharing with me, on that occasion, the fond memories of Feza G\"{u}rsey and Larry Biedenharn, to whom I am indebted for triggering my interest in  the exceptional Jordan and Lie structures, so relevant in physics. I also wish to thank Alessio Marrani, Michael Rios and Klee Irwin for dragging me out of silence and stimulating me to go back to research. I thank Klee for supporting my research in quantum gravity: his initiative has gathered at QGR -Quantum Gravity Research- one of the largest group of scientists entirely devoted to the solution of the biggest mystery in the Pandora's vase.

\end{document}